\documentstyle[11pt,psfig,vsopsymp]{article}
\begin{document}


\markboth{\hfill Jones et al.}{Space VLBI at Low Frequencies \hfill}


\centerline{\LARGE\bf 
   Space VLBI at Low Frequencies  
}\bigskip

\centerline{\sc 
D.~L.~Jones$^1$, R.~Allen$^2$, J.~Basart$^3$, T.~Bastian$^4$,} 
\centerline{\sc 
W.~Blume$^1$, J.-L.~Bougeret$^5$, B.~Dennison$^6$, M.~Desch$^7$,}
\centerline{\sc 
K.~Dwarakanath$^8$, W.~Erickson$^9$, W.~Farrell$^7$, D.~Finley$^4$,}
\centerline{\sc 
N.~Gopalswamy$^7$, R.~Howard$^{10}$, M.~Kaiser$^7$, N.~Kassim$^{11}$,}
\centerline{\sc 
T.~Kuiper$^1$, R.~MacDowall$^7$, M.~Mahoney$^1$, R.~Perley$^4$,}
\centerline{\sc 
R.~Preston$^1$, M.~Reiner$^7$, P.~Rodriguez$^{11}$, R.~Stone$^7$,}
\centerline{\sc 
S.~Unwin$^1$, K.~Weiler$^{11}$, G.~Woan$^{12}$ \& R.~Woo$^1$  
}\medskip

\centerline{\it 
$^1$ Jet Propulsion Laboratory, California Institute of Technology, USA
}
\centerline{\it 
$^2$ Space Telescope Science Institute, USA 
}
\centerline{\it 
$^3$ Iowa State University, USA 
}
\centerline{\it 
$^4$ National Radio Astronomy Observatory, USA 
}
\centerline{\it 
$^5$ Observatoire de Paris, France 
}
\centerline{\it 
$^6$ Virginia Polytechnic Institute, USA 
}
\centerline{\it 
$^7$ Goddard Space Flight Center, USA 
}
\centerline{\it 
$^8$ Raman Research Institute, India 
}
\centerline{\it 
$^9$ University of Maryland, USA, \& University of Tasmania, Australia
}
\centerline{\it 
$^{10}$ Orbital Sciences Corp., USA 
}
\centerline{\it 
$^{11}$ Naval Research Laboratory, USA 
}
\centerline{\it 
$^{12}$ University of Glasgow, UK 
}

\begin{abstract} \noindent 
At sufficiently low frequencies, no ground-based radio array will  
be able to produce high resolution images while looking through 
the ionosphere.  A space-based array will be needed to
explore the objects and processes which dominate the
sky at the lowest radio frequencies.   
An imaging radio interferometer based on a large number of
small, inexpensive satellites would be able to track solar
radio bursts associated with coronal mass ejections out to
the distance of Earth, determine the frequency and duration
of early epochs of nonthermal activity in galaxies, and
provide unique information about the interstellar medium.
This would be a ``space-space" VLBI mission, as only baselines 
between satellites would be used.  
Angular resolution would be limited only by interstellar and
interplanetary scattering.  
\end{abstract}


\sources{ }

\section{Introduction}

Ground-based radio interferometers 
are able to produce images of the sky at frequencies down to  
a few tens of MHz.  Some important scientific goals, however,
require imaging at even lower frequencies.  Absorption and 
refraction by the ionosphere prevents imaging from the ground
at frequencies of a few MHz and lower, so an interferometer 
array composed of inexpensive satellites will be needed.   
Suitable locations for a space-based array include very high
Earth orbits, halo orbits about the Sun-Earth Lagrange points, 
Earth-trailing heliocentric orbits, the far side of the
Moon, and (perhaps) lunar orbit.  The optimal choice depends 
on financial considerations 
and the unavoidable tradeoff between a benign environment in
which to maintain a multi-satellite array and the difficulty
of getting enough data from the array to Earth.

\section{Science Goals} 

What unique science can be done only at frequencies below $\sim10$
MHz?  There are two general areas where very low frequency 
observations are critical:  First, sources of emission which
are intrinsically limited to low frequencies (e.g., plasma 
oscillations and electron cyclotron masers), and second, observations
of strongly frequency-dependent absorption (e.g., free-free absorption 
by diffuse ionized interstellar hydrogen).  Type II radio bursts 
from interplanetary shocks driven by coronal mass ejections 
provide a good example of the first case.  These intrinsically
narrow-band emissions decrease in frequency as the shock propagates
farther from the Sun into regions of lower plasma density.  In
order to image and track type II bursts as they approach 1 AU
from the Sun, observations at frequencies below 1 MHz are 
necessary.  This would allow us to predict the 
arrival at Earth of coronal mass ejections, which can trigger 
severe geomagnetic storms. 
If located far enough from Earth, a low frequency array would
also be able to image Earth's magnetosphere from the outside 
and observe how it changes in response to solar disturbances. 

A sensitive map of the radio sky with arcminute angular resolution
at a few MHz would be especially effective at detecting coherent
emission from disks, jets, and possibly gas giant planets orbiting
close to nearby stars.  
Most coherent processes have sharp upper-frequency limits, and
can only be detected at low frequencies.  

All-sky surveys at low frequencies would map the galactic distribution  
of low energy cosmic ray electrons and would likely discover large 
numbers of high redshift galaxies, ``fossil" radio lobes, and 
large-scale interstellar shocks and shells from old galactic 
supernovae and $\gamma$-ray bursts. 
In addition, diffuse ionized hydrogen could
be detected via its absorption of radiation from extragalactic 
radio sources across the sky.  These observations would complement
H$\alpha$ emission maps, which predict large variations in free-free
optical depth on angular scales of a few degrees. 

\section{Requirements for a Low Frequency Array in Space} 

\begin{center}
\begin{figure}[ht]
\vspace{121mm}
\includegraphics{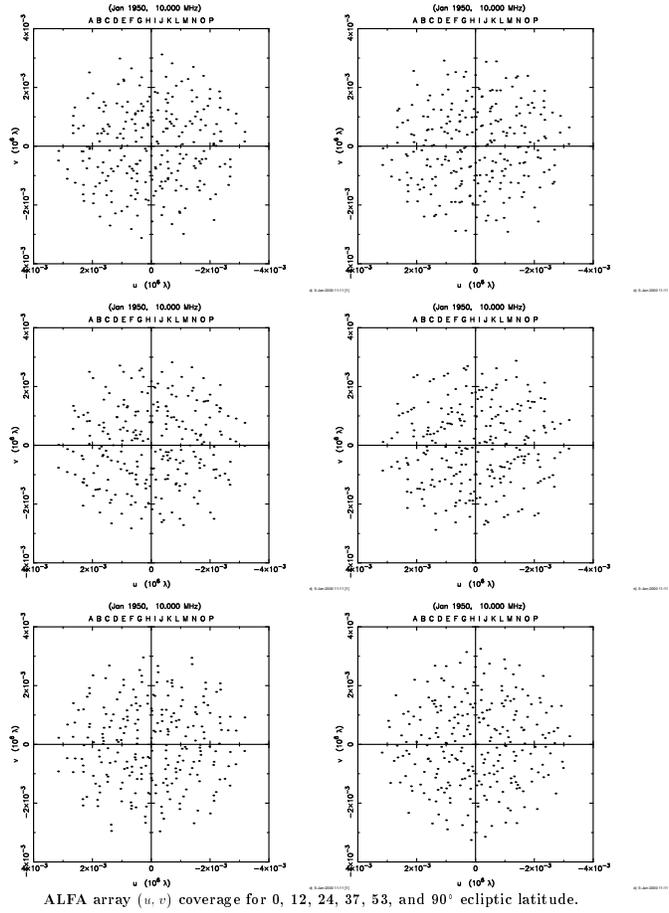}
\caption{Snapshot (instantaneous) $u,v$ coverage provided 
by a 16 satellite spherical array over a wide range of  
directions simultaneously.}
\label{Fig1}
\end{figure}
\end{center}

Any space-based array for very low frequency imaging will need
to meet three fundamental requirements:  1) the array must be
located far enough from Earth to avoid terrestrial interference
and the extended ionosphere, 2) there must be a large enough 
number of individual antennas in the array to produce dense,
uniform ($u, v$) coverage in all directions 
simultaneously, and 3) the observing bandwidth must be sufficient
to provide useful sensitivity for short snapshot observations. 
The second and third requirements result from the nearly 
omnidirectional nature of reasonably sized antennas at very
low frequencies.  Strong variable radio sources
anywhere on the sky will affect the observed total power levels,
and unless such sources are imaged on short time scales their
time-variable sidelobes will limit the dynamic range 
of observations in other directions.   
Simulations show that a minimum of 12 satellites will be
needed, with at least 16 satellites preferred 
(see Figure \ref{Fig1} on the next page). 

The maximum useful baseline length is set by interstellar and
interplanetary scattering.  These effects are proportional to
$\nu^{-2}$ and thus are much stronger at low frequencies.  
For frequencies of a few MHz, maximum baselines of a few 
hundred km are appropriate.  The
degree of scattering at any given frequency is a strong
function of direction on the sky.  Consequently, is it 
important to have a wide range of baseline lengths in the
array.  Short projected baselines are needed in any case to
allow angularly large structures in solar radio bursts and
the galactic synchrotron background to be imaged. 

Imaging the entire sky is a daunting task, but it can be
made tractable by dividing the sky into $\sim10^{3}$ 
fields of view and relying on parallel processing.  Each
field will require only a 16-bit Fourier transform in the
radial direction to account for sky curvature, and $\approx 100$
deconvolving beams (see Frail \etal\ 1994).
During deconvolution, model components from all fields must
be subtracted from the full 3-D visibility data set to
remove sidelobes from other fields during the next iteration
of residual image production.  The total computing rate 
required is large by current standards, but will be readily
available within a few years.  

\acknowledgements
Part of this work was carried out at the Jet Propulsion Laboratory,
California Institute of Technology, under contract with the US
National Aeronautics and Space Administration.


\begin{references}

\newref Frail,~D., Kassim,~N. \& Weiler,~K. 1994, \aj, {\bf 107},
1120
\end{references}
\end{document}